# Transmission measurement at 10.6 μm of $Te_2As_3Se_5$ rib-waveguides on $As_2S_3$ substrate

C. Vigreux-Bercovici [a)], E. Bonhomme and A. Pradel
*Laboratoire de Physico-chimie de la Matière Condensée, Institut Gehrardt, UMR 5617, Université Montpellier II, Place Eugène Bataillon, 34095 Montpellier cedex 5, France*

J.-E. Broquin
*Institut de Microélectronique, Electromagnétisme et Photonique, ENSERG, 23 av. des Martyrs, BP 257, 38016 Grenoble cedex 1, France*

L. Labadie and P. Kern
*Laboratoire d'Astrophysique de l'Observatoire de Grenoble, Observatoire de Grenoble, BP 53, 38041Grenoble cedex 9, France*

**Abstract**

The feasibility of chalcogenide rib waveguides working at $\lambda = 10.6\mu m$ has been demonstrated. The waveguides comprised a several microns thick $Te_2As_3Se_5$ film deposited by thermal evaporation on a polished $As_2S_3$ glass substrate and further etched by physical etching in Ar or $CF_4/O_2$ atmosphere. Output images at 10.6μm and some propagation losses roughly estimated at 10dB/cm proved that the obtained structures behaved as channel waveguides with a good lateral confinement of the light. The work opens the doors to the realisation of components able to work in the mid and thermal infrared up to 20 μm and even more.

**Text**

The development of single-mode integrated optics components being able to work in the whole range from 1 to 20 microns is of major interest for specific applications related to detection in the mid and thermal-infrared spectral domain : the detection and identification of pollutant gases for environmental control or the direct detection of exoplanetary systems (ESA Darwin Mission [1]), for example. To date, single-mode integrated optics is limited to the near infrared windows H [1.50 – 1.80 μm] and K [2.01 – 2.42 μm] due to the silica transmission window. One of the promising possibilities to extend the integrated optics concept to the mid-infrared is to use chalcogenide glasses which transparency in the infrared is well-known. Some channel waveguides mainly based on arsenic sulphide $As_2S_3$ or arsenic selenide $As_2Se_3$ were already realised [2, 3 for example]. In reference [2], rib waveguides made by a physical or reactive ion etching of $As_2S_3$ were optically characterised at $\lambda = 1.5$ microns. In reference [3] channel waveguides were obtained by direct laser writing of the $As_2Se_3$ core layer deposited on an $As_2S_3$ cladding layer and were characterised at $\lambda = 8.5$ microns. In both examples, the substrates were crystalline silicon substrates ($Si/SiO_2$) and the working domain of the structures was limited at 12 microns, due to the opacity of $As_2S_3$ at longer wavelengths. In order to extend the working domain of such waveguides, the only way is the replacement of the sulphide and selenide films by telluride ones. The number of papers dealing with telluride glasses in bulk form has been growing up in the past few years [6-8] but only a few of them concern films [8-10]. Moreover in previous works on telluride films, the film thickness was lower than 1 micron [8-10] while the film thickness will have to be typically from 4 to 15 microns for use in the mid-IR. In fact, the further the waveguides will be used in the IR, the thicker the film will have to be. For similar reasons, the films will have to be etched on variable depths, typically from 1 to 10 microns. These dimensions (film thickness and rib depth) are much bigger than the ones usually used for the telecommunication wavelengths and previous tests to reach such characteristics are scarce in the literature [2;11].

The present work was dedicated to demonstrate the feasibility of channel waveguides based on telluride films and able to work in the IR (10.6 microns). In this first stage, only multimode waveguides were considered and the guiding film alone was made of telluride. Our choice was to realise all chalcogenide rib waveguides with a chalcogenide substrate on the top of which was deposited a telluride thick film. Such a choice helped in avoiding the stacking of several thick films and potential problems of mechanical stress and lack of adhesion. The substrates used in the work were commercially available $As_2S_3$ plates of 3 mm in thickness. The telluride composition was $Te_2As_3Se_5$ since it is well known in its bulk form [4; 5]. First, the fabrication of the rib waveguides required the deposition of the telluride glass in film form while respecting its composition and refractive index. The geometry of the film had then to be modified by etching, in order to achieve the lateral confinement of the light.

Two deposition methods were tested to deposit $Te_2As_3Se_5$ as a thick film: RF-sputtering of home-made telluride glass target and thermal evaporation of home-made telluride powders. The main drawback of the RF sputtering processes is the low deposition rate, which makes the deposition of thick films unrealistic. The low density of the deposited films is a second drawback of the technique [12]. Therefore, the thermal evaporation process was selected since it allows obtaining dense and thick films of $Te_{20.6\pm0.3}As_{30.7\pm0.3}Se_{48.7\pm0.3}$ composition with very high deposition rates (about 1 $\mu m.min^{-1}$) [12]. A value of 2.821±0.005 for the refractive index of the films at $\lambda$ = 10.6 μm was measured thanks to m-lines measurements [13]. Note that the refractive index of the commercial $As_2S_3$ glass used as a substrate is 2.38 at $\lambda$ = 10.6 μm.

Two types of processes were tested to etch the obtained films: physical etching in an argon atmosphere and reactive ion etching in a $CF_4/O_2$ atmosphere [11]. The first method allowed obtaining ribs with heights up to 2.7 microns, i.e. close to the thickness of the resin bands obtained by lithography on the film surface. Such a result is consistent with the fact that the argon etching is not selective. Reactive ion etching in a $CF_4/O_2$ atmosphere allowed increasing the film etching rate and making the etching selective. Fig.2 shows the etching rates of the film and the resin versus the oxygen amount (with constant total gas flow). Selectivity is at its maximum at about 15 % $O_2$ when the positive effect of oxygen in increasing the appearance of $F^-$ that reacts with the layer elements is stronger than its negative effect of producing a passivated oxidized layer. Such a result is in agreement with data from reference [2]. In these conditions, thanks to high rate and selectivity, it was possible to obtain quasi vertical (angle higher than 70 °) ribs with height of 5 microns.

Before performing the optical tests and in order to obtain an efficient injection of the light, the input and output facets of the rib waveguides have been polished with alumina powders with particle sizes of 5, 3, 1 and 0.3 microns, successively.

A preliminary optical near-field characterisation at 1.5 μm has been carried out in order to check the injection efficiency and to estimate the waveguiding properties of the samples. The light was coupled into the waveguides by a single mode optical fibre. The output intensity was focused by means of an X25 microscope objective on an IR camera. Once qualified at λ = 1.55 µm, the samples have been characterised on an optical bench dedicated to measurements in the mid-infrared at a wavelength of 10.6 µm. The main data obtained on a typical waveguide "$Te_2As_3Se_5$ film/$As_2S_3$ bulk substrate" are shown on figure 2. The waveguide was fabricated by depositing a $Te_2As_3Se_5$ film of 4 microns in thickness on a polished $As_2S_3$ substrate. 15 µm-large bands of resin were created on the film surface by traditional photolithographic method. Then, the film was physically etched in an argon atmosphere creating a 1.9µm high rib. A SEM image of the as-obtained rib is shown in figure 2a along with its profile as measured by profilometry. During the optical characterization at λ = 10.6 µm, the injection spot was either centred on the ribs or on the contrary, de-centred from a rib waveguide and rather in-between two ribs or below a rib as shown in figure 2b. Two different output signals were then observed depending on the injection spot position [14]. In the first case, the output signal was characteristic of a good lateral confinement of the light (Fig.2c). In the second case, the output signal was characteristic of a planar waveguide (Fig.2d). This behaviour is typical of rib waveguides which can propagate the unguided radiated field over quite a long length when the distance between the ribs (apart by 385 µm) allows the propagation of a planar waveguide mode. The transmission of the 1cm long component was evaluated at 1.01 % ± 0.03 %, which corresponds to insertion losses of roughly 20dB. Since these losses include coupling and propagation losses. Such a weak transmission is neither surprising, neither prohibitive. As a matter of fact, the present waveguides are not optimized ones. Indeed, a large difference exists between the refractive indices of the telluride core (2.821) and the substrate, i.e. $As_2S_3$ (2.38) and air. Therefore losses due to reflection at both interfaces can be estimated using Fresnel's reflection coefficient to be 1.1dB per facet. Moreover, an estimation of the coupling losses can be derived from the calculation of the fundamental mode waveguide dimensions, which are 14 µm by 4 µm (at 1/e) and by computing the overlap of this mode with the input laser spot, which diameter has been measured to be 30.2 µm. With all these data, the coupling losses due to mode shape mismatch have been estimated to 7 dB. Thus, almost 10 dB of the insertion losses are due to coupling issues that can be solved by reducing the refractive index contrast between the substrate and the core of the waveguide. Indeed, a change of the sulphide substrate for a telluride one and the addition of a chalcogenide film as a superstrate is feasible. Note that the deposition of a telluride film on the top of a telluride bulk glass has recently been successfully carried out [15]. In fact, reducing the refractive index contrast will also help improving the modicity of the waveguides since the obtained waveguides are not single-mode ones and sustain three propagation modes though these latter are not resolved in the output signal shown in figure 2c and can not be directly observed.

    The feasibility of rib waveguides based on telluride thick films and able to work at 10.6 µm has therefore been demonstrated. Total insertion losses of a 1 cm long sample are 20 dB, which can be splitted in 10 dB for coupling losses and 10 dB for propagation losses. It opens doors for the realisation of components for (thermal) infrared integrated optics. As a matter of fact, a judicious choice of telluride compositions for both the substrate and the film will help in developing components able to work up to 20 µm and even more.

**Acknowledgements**
    We thank the European Space Agency for its financial support within the contract "Integrated Optics for Darwin" and Alcatel Alenia Space for its interest in the work. We also thank Yves Moreau and Raphael K. Kribichof from the CEM2 in Montpellier (France) for helping in the characterisation of the waveguides at 1.5 microns.

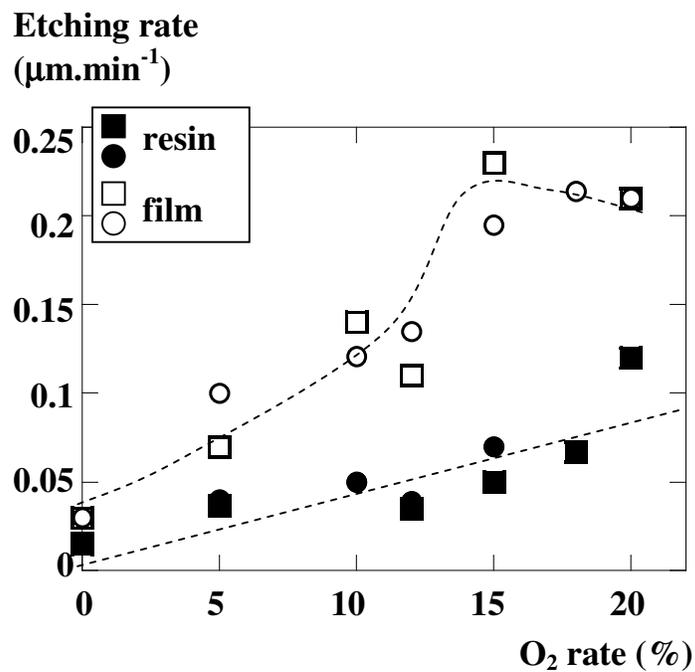

FIG. 1. Reactive ion etching rates of the film and the resin versus the $O_2$ percentage for two different sets of data (the lines are guides for the eyes). A total gas flow of $CF_4 + O_2$ of 75 sccm was applied. Etching was performed in a SEMCO Eng. GIGAX 50 sputtering set-up, applying a RF power of 20 W and an ECR power of 200 W.

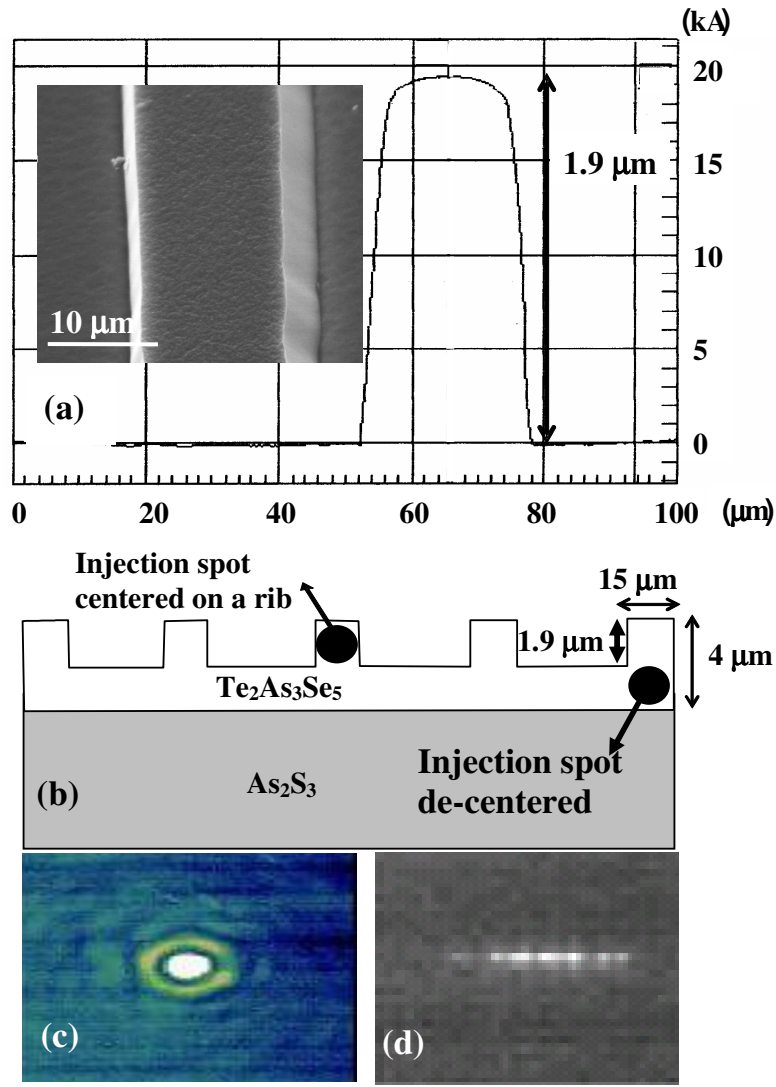

FIG.2. (a) SEM image and rib profile of a rib waveguide "$Te_2As_3Se_5$ film/$As_2S_3$ bulk substrate", (b) sketch of the rib-waveguide showing the position of the injection spot, (c) output image at 10.6 microns when the injection spot was centred on a rib, (d) output image at 10.6 microns when the injection spot was de-centred from the ribs.